\documentclass[twocolumn]{aa}
\usepackage{graphicx}
\usepackage{txfonts}
\begin{document}

    \title{The influence of jet geometry on light curves \\
           and spectra of GRB afterglows}

    \author{A.G.~Tolstov
          \inst{1}\inst{2}
    }

    \offprints{A.G.~Tolstov}

    \institute{Institute for Theoretical and Experimantal Physics (ITEP),
               Bolshaya Cheremushkinskaya, 25, 117218 Moscow, Russia\\
               \email{tolstov@mail.itep.ru}
        \and
               Max Plank Institute for Astrophysics(MPA),
               Karl-Schwarzschild-Str., 1, 85741 Garching, Germany\\
               \email{tolstov@mpa-garching.mpg.de}
    }

    \date{Received May 26, 2004; accepted November 27, 2004}

    \abstract{We have performed detailed calculations of spectra and
              light curves of GRB afterglows assuming that the observed GRBs can
              have  a jet geometry. The calculations are based on an expanding
              relativistic shock GRB afterglow model where the afterglow is the
              result of synchrotron radiation of relativistic electrons with
              power-law energy distribution at the front of external shock being
              decelerated in a circumstellar medium. To determine the intensity
              on the radiation surface we solve numerically the full time-,
              angle-, and frequency-dependent special relativistic transfer
              equation in the comoving frame using the method of long
              characteristics.

    }

\maketitle
%
\section{Introduction}

    At the present time we know that gamma-ray bursts (GRBs) are
    explosive phenomena at cosmological distances. If the emission is
    isotropic, estimations based on observations give us the values of
    released energy up to $E_0 \sim 3.4 \times 10^{54}$ ergs for
    GRB990123, that exceeds the rest energy of a solar mass star
    (Kulkarni et al. \cite{Kulkarni}). To reduce this large amount of
    energy it can be supposed that the GRB emission is highly
    collimated. May be, the better evidence for jet structure is the
    achromatic break in light cirves (Sari \cite{Sari99a}) of the light curves
    seen in many afterglows, e.g. GRB990123 (Kulkarni et al.
    \cite{Kulkarni}) and GRB990510 (Harrison \cite{Harrison}, Stanek
    \cite{Stanek}). And, finally, spherical symmetry conflicts with
    linear polarization (Sari \cite{Sari99b})
    observed for a few afterglows (Covino \cite{Covino}, Wijers
    \cite{Wijers}).

    Generally, a GRB jet can display an angular structure and can be
    seen by observer at wide range of viewing angles from the jet axis
    (Wei \& Jin \cite{Wei}, Granot \cite{Granot03}). For now, however, we
    consider a jet with uniform angular structure
    taking into account the effect of equal-arrival-time surface at
    different angles of observation and show which changes in GRB
    afterglow are produced in transition from spherical symmetry to
    jet geometry.

    The evolution of the jet and the light curves has been widely
    investigated (Panaitescu \cite{Panaitescu}, Kumar \cite{Kumar}),
    including lateral jet expansion (Salmonson \cite{Salmonson}),
    investigation of the 'stuctured' jet (Granot \cite{Granot03}), 3D
    numerical simulations of the jet dynamics (Canizzo \cite{Canizzo})
    and different angles of observation (Granot \cite{Granot02}). All
    these works have not solve accurately transfer equation for the
    resulting light curves calculations and are based either on simple
    expressions for local emissivity or focus on
    power-law branch of the spectrum between the break frequencies or some other simple
    assumptions on the characteristics of radiation field.

    In this paper we present a detailed calculation of spectra and
    light curves with preliminary
    numerical solution of special relativistic transfer equation in
    the comoving frame. We will show that the exact calculation of intensity,
    depending from the angle to the surface normal, can have an essential
    influence on the form of the spectra visible to the
    observer. The exact determination of equitemporal surfaces
    which is important for explanation of observed luminosity in
    GRBs and comprehension of their spectral properties
    (Bianco \& Ruffini \cite{Bianco}) is also
    taken into account in our calculations.

    The transfer equation needs to be solved for the accurate
    calculation of intensity on the surface of radiating structure by
    integrating the emission along the characteristic through the
    structure and the following fluxes calculations. The calculation
    is based on the model where the jet is cut from a spherically
    symmetric flow. We add two parameters for taking into account a
    radial jet structure and different values of observer
     viewing
    angles. In the next section we discuss our model
    in more detail,
    in section 3 we calculate the emission for different values of the
    parameters and finally we present some discussions and
    conclusions.

    \begin{figure}
    \centering
    \includegraphics[width=88mm]{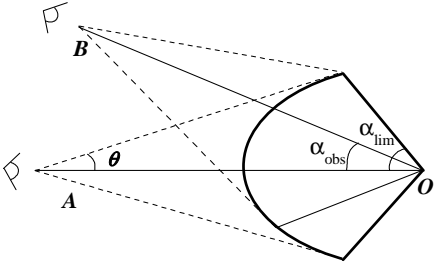}
    \caption{The part of quasi-ellipsoid from which the photons reach
             a remote observer. The explosion center is located at the vertex
             of the jet opening angle $\alpha_{\rm lim}$. The jet is observed
             under the angle $\alpha_{\rm obs}$ between the jet symmetry axis
             and the direction towards the observer.
    }
    \label{Fig1}
    \end{figure}


\section{Physical Model}

    Our jet geometry investigation is based on the numerical solution
    of the problem in the spherically-symmetric case. To take into
    account jet geometry first we fix the direction towards the
    observer in the case of spherical symmetry. Because of the high
    shock velocity light at a certain time reaches the observer from
    the ellipsoidal structure, the part of which can be seen in
    Fig.~\ref{Fig1}. To construct the jet we cut the cone with axis
    being the direction towards the observer $A$ and the angle
    $\alpha_{\rm lim}$, forming jet opening angle. The sight of
    observer $A$ in this case coincides with jet axis. To consider the
    jet under different values of viewing angle we add angle
    $\alpha_{\rm obs}$ between the observer $B$ and the jet axis. This
    approach gives us the possibility to consider some jet effects
    without using specific hydrodynamics code and more complicated
    transfer equation.

    Let us consider the model we have used for numerical calculations
    of spectra and light curves in the case of spherical symmetry
    (Tolstov \& Blinnikov \cite{Tolstov}).


    In general, the transfer and hydrodynamic equations constitute a
    combined system of equations. In our problem, however, we solve
    them separately. To determine the variables of the medium we use a
    self-similar solution for a relativistic shock for an
    ultrarelativistic gas in the case of the conservation of total
    shell energy (Blandford \& McKee \cite{Blandford}). The solution describes
    the explosion with a fixed amount of energy $E_0$ and propagation
    of a relativistic shock through a uniform cold medium.
    \[
        p=\frac{1}{3}e=\frac{2}{3}w_1\Gamma^2 \chi^{-17/12}; \quad
        \gamma^2=\frac{1}{2}\Gamma^2 \chi^{-1}; \quad
        n\gamma=2n_1\Gamma^2 \chi^{-7/4} ,
    \]
    where
    \[
        \Gamma^2\propto t^{-3}; \quad
        \chi=[1+8\Gamma^2](1-r/t).
    \]
    Above, $\Gamma$ is the Lorentz factor of the shock front, $\gamma, p, e, n$ -
    the Lorentz factor, pressure, energy and density of the shocked fluid, measured
    in the local rest frame of fluid, respectively, and $n_1$ is density of the external medium.

    For accurate calculation we should know also the electron energy
    spectrum and the magnetic field strength. Here for local
    emissivity calculation we use the conventional assumptions from
    relativistic electrons (e.g. Sari \cite{Sari99c}).
    We assume being based on standard fireball shock model (Zhang \cite{Zhang})
    that the electrons have a power-law distribution and that their total
    energy behind the shock front accounts for $\epsilon_e$ of the
    internal energy:
    \[
        N(\gamma_e)=K_0\gamma_e^{-p}; \quad \gamma_e \ge
        \gamma_{min,0}=\frac{ \epsilon_e e_0}{n_0 m_e c^2},
    \]
    where $m_e$ is the electron rest mass and $K_0=(p-1)n_0\gamma_{min,0}^{p-1}$.

    The magnetic field is parameterized by the quantity $\epsilon_B$,
    which is equal to the fraction of the internal energy contained in
    the magnetic field $ B^2=8\pi\epsilon_B e .$ The magnetic field is
    randomly oriented and decreases with time due to adiabatic
    expansion of the shell.

    As the electrons pass through the shock they begin to lose the
    energy through adiabatic cooling determined by the solution of
    Blandford \& McKee (\cite{Blandford}). This process is well
    described in detail by Granot \& Sari (\cite{Granot02}).

    Here we present only the basic formulas for synchrotron
    radiation used in our calculation.
    The spectral power of a single electron averaged over the pitch angle is:
    \[
        P(\omega)=\frac{3^{5/2}}{8\pi}\frac{P_{sy}}{\omega_0}
        F\Big(\frac{\omega}{\omega_c}\Big) ,
    \]
    where
    \[
        P_{sy}=\frac{1}{6\pi}\sigma_{Th}cB^2(\gamma_e^2-1);\quad
        \omega_c=\frac{3\pi}{8}\frac{eB}{m_ec}\gamma_e^2 ,
    \]
    and $F(u)$ - a standard synchrotron radiation function (Rybicki
    \cite{Rybicki}). The synchrotron absorption coefficient is
    specified by the formula:
    \[
        \chi = \frac{1}{8\pi m_e\nu^2}
        \int_{\gamma_{\min}}^{\gamma_{\max}}d\gamma\frac{N(\gamma)}{\gamma^2}\frac{d}
        {d\gamma}\Big(\gamma^2P(\omega,\gamma)\Big)
    \]

    To determine the intensity on the radiation surface we solve
    numerically the full time-, angle-, and frequency-dependent
    relativistic transfer equation in the comoving frame using the
    method of long characteristics up to the values of Lorentz-factor
    $\gamma \sim 1000 $.

    After the calculation of intensity the flux can be determined:
    \[
        F_{0,t_{\rm obs}}=\frac{2\pi}{D^2} \int_{\mu_{0,min}}^{1} \mu R^2
        I(r(\mu_0),\nu_0\Big(\frac{\nu}{\nu_0}\Big),\cos\delta(\cos\delta_0))
        \Big(\frac{\nu_0}{\nu}\Big)^3 d\mu_0 ,
    \]
    where subscript $_{0}$ is related to the observer frame,
    $D$ is the distance to observer, $p=R/D$. $\mu_{min}$ -
    cosine of the maximum angle visible to the observer.

    To sum up, the observer afterglow spectra and light curves depend
    on the the hydrodynamic evolution, the radiation processes, the
    distance to the observer and the two parameters we have used for
    taking into account jet structure: the jet opening angle
    $\alpha_{\rm lim}$ and the viewing angle $\alpha_{\rm obs}$.

    In our calculation for solving the problem in the case of
    spherical symmetry we have used the following parameters: $E_0 =
    10^{53}$ ergs, $n_1 = 1$ cm$^{-3}$, $\epsilon_e=0.5$,
    $\epsilon_B=0.1$, $p=2.5$, $D=10^{27}$ cm. In the next section we
    consider the results of the jet geometry influence on spectra and
    light curves varying $\alpha_{\rm lim}$ and $\alpha_{\rm obs}$.

    \begin{figure}
    \centering
    \includegraphics[width=88mm]{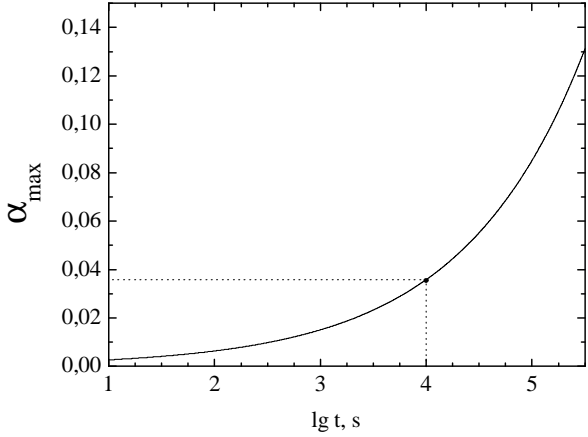}
    \caption{The angle visible to observer from emitting structure
             increases with time.  Time is measured in the observer frame of
             reference. For the $t=10^4$ s it is shown that there is no reasons
             to increase the jet opening angle more that $0.037$.
    }
    \label{Fig2}
    \end{figure}

    \begin{figure}
    \centering
    \includegraphics[width=88mm]{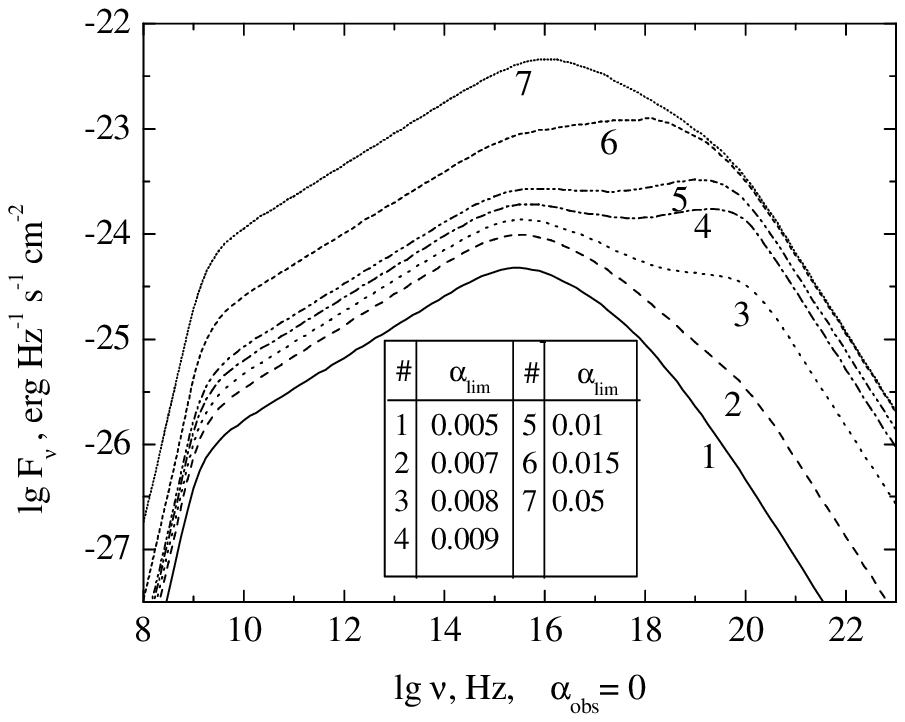}
    \caption{Instantaneous afterglow spectra at time $t=10^4$ s and
             observational angle $\alpha_{\rm obs}=0$ for different values of
             limitation angle $\alpha_{\rm lim}$.
    }
    \label{Fig3}
    \end{figure}

    \begin{figure}
    \centering
    \includegraphics[width=88mm]{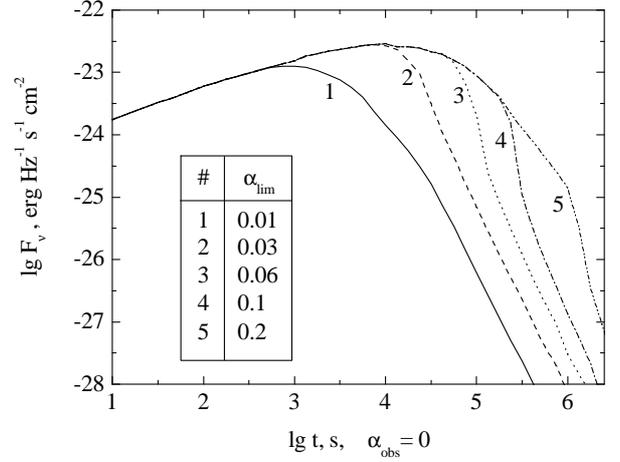}
    \caption{Afterglow light curves at frequency $\nu=5\times 10^{14}$
             Hz and observational angle $\alpha_{\rm obs}=0$ for different
             values of limitation angle $\alpha_{\rm lim}$.
    }
    \label{Fig4}
    \end{figure}


\section{Results of the numerical calculation}

    As the shell from which the light reaches the observer moves
    towards the observer, the angle of the structure visible to the
    observer increases with time. This dependence is presented in
    Fig.~\ref{Fig2}, where $\alpha$ is the angle between the direction
    to the observer and the line connecting the center of the symmetry
    with the point on the surface that is still visible to the
    observer. There is the
    following relationship between the real observation angle $\theta$ and $\alpha$ has :
    \[
        \cos\theta=\frac{1-\mu p(\mu)}{[1+p^2(\mu)-2p(\mu)\mu]^{1/2}},
    \]
    where $\mu=\cos\alpha$, $p=R(t_{\rm obs})/D$, $t_{\rm obs}=t_{\rm
    obs}(\mu,p,D)$, $D$ -
    distance from the center of the burst to the observer, and time in
    the burst frame of reference $t$ is connected with the time in the
    observer frame of reference $t_{\rm obs}$ by the formula:
    \begin{eqnarray}
        && t_{\rm obs}=t+\frac{D(1-[1+p^2(\mu)-2p(\mu)\mu]^{1/2})+R_0}{c}\approx
        \nonumber \\
        && \approx t+ \frac{R_0-R(t_{\rm obs})}{c} \quad (p \ll 1) \nonumber
    \end{eqnarray}
    if we suppose that the initial time of observation
    corresponds to the initial time of the burst $(t_0 =
    t_{0,obs}+D/c)$ and initial radius of the burst structure is
    $R_0$.

    Now if we fix the time of observation by the value, say, of
    $t_{\rm obs}^{'}=10^4$ c, there is no reason to increase the value
    of limitation angle in our jet structure for more than
    $\alpha_{\rm lim}^{'}=0.037$ (Fig.~\ref{Fig2}), because this does
    not produce any effect on the resulting spectra and light curves
    as if they are considered from the non-limited structure.

    In Fig.~\ref{Fig3} and Fig.~\ref{Fig4} we can see the calculated
    spectra and light curves at zero observational angle and at
    different values of limitation angle.

    The changed form of the spectra, having at some values of
    limitation angle two peak fluxes, is the consequence of the ring
    intensity structure on the radiative surface (Tolstov \&
    Blinnikov \cite{Tolstov}).

    If we look at the ring structure we see that
    than more the light frequency than closer the maximum of brightness to the edge of the image.

    The flux from the observed
    image can be calculated by the formula
    \[
        F_{\nu}=2\pi\int_{\cos\theta_{max}}^{1}I(\cos\theta,r,\nu)\cos\theta
        d\cos\theta
    \]
    where $\theta$ is the angle between the point on the radiating
    surface and the center of the structure (see Fig.~\ref{Fig1}),
    $\theta_{max}$ corresponds to the edge of the image. In the
    absence of jet limitation angle the flux is monotonically
    decrease in increasing light frequency at the right part of
    the spectrum. In the presence of limitation angle
    some maximums of intensity at lower frequencies are excluded
    from the flux integral and it gives at higher frequencies flux value
    compatible to that one at lower frequencies.

    The light curves do not show this effect just having ''jet
    breaks'' due to the limitation angle. This results from decreasing
    of the radiation arrived to the observer from the shock limited by
    $\alpha_{lim}$. Larger the value of $\alpha_{lim}$, less radiation
    at some frequencies gets towards the observer. Of order of days at
    $\alpha_{lim}=0.2$ we can see the break typical for some observed
    optical afterglows (Zhang \& Meszaros \cite{Zhang}).

\begin{figure}
    \centering
    \includegraphics[width=88mm]{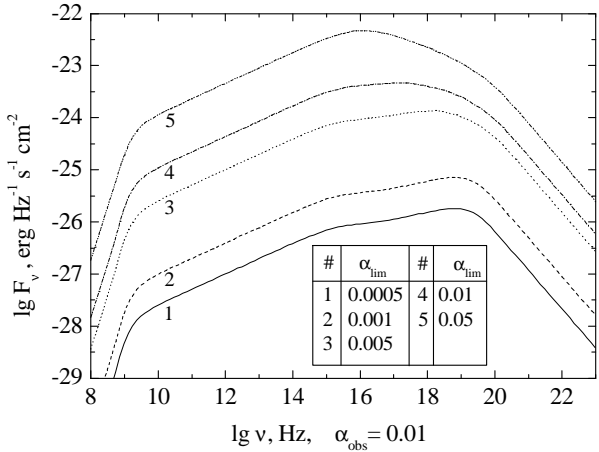}
    \caption{Instantaneous afterglow spectra at time $t=10^4$ s and
             observational angle $\alpha_{\rm obs}=0.01$ for different values
             of limitation angle $\alpha_{\rm lim}$.
    }
    \label{Fig5}
    \end{figure}

\begin{figure}
    \centering
    \includegraphics[width=88mm]{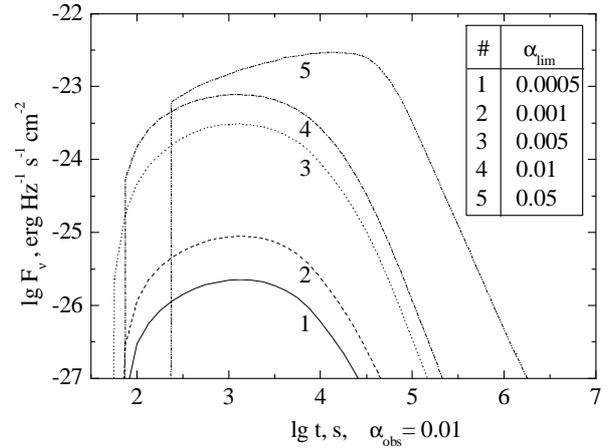}
    \caption{Afterglow light curves at frequency $\nu=5\times 10^{14}$
             Hz and observational angle $\alpha_{\rm obs}=0.01$ for different
             values of limitation angle $\alpha_{\rm lim}$.
    }
    \label{Fig6}
    \end{figure}

    In Fig.~\ref{Fig5} and Fig.~\ref{Fig6} we present the calculated
    spectra and light curves at the observational angle $\alpha_{\rm
    obs}=0.01$. At small values of limitation angles we can see also
    the changes in spectra but for the light curves we did not have
    this effect. Some of the presented light curves are cut at early
    moments of time. This is the consequence of solving transfer
    equation only up to the Lorentz-factor value $\gamma = 1000$. At
    larger value of Lorentz-factor we suppose that the matter is not
    radiative and if at $\alpha_{\rm obs}=0$ this effect is not
    revealed that at $\alpha_{\rm obs}=0.01$ some of the
    characteristics start at Lorentz-factor value $\gamma \ge 1000$.

\section{Conclusions}

    It is widely believed that GRBs are born in jet geometry. In this
    case the resulting afterglow radiation becomes highly collimated.
    The numerical calculations of light curves in these models (Granot
    \cite{Granot03}, Salmonson \cite{Salmonson} )
    are based on some assumptions for intensity on the propagating
    shock front. The shape of the local spectral
    emissivity is approximated as a broken power-law with some typical
    breaks corresponding to synchrotron radiation. As we can see from
    our results the spectra can have some peculiarities and the shapes
    different from power-law as in direct view to the jet as at some angle
    to the jet axis.

    Of course, our consideration does not take into account some
    effects of jet model and the exact numerical calculations should
    be at least two-dimensional to allow lateral expansion and angular
    structure of the jet.

    Nevertheless we would like to point out that the accurate
    calculation of intensity using relativistic transfer equation can
    have an influence on the shape of spectra and light curves of GRB
    afterglow. Constructing more precise model using the exact
    numerical calculations can help to explain the peculiarities of
    GRB afterglow and shed some light on the nature of GRB phenomenon.

\begin{acknowledgements}
    I am grateful to S.I. Blinnikov for posing the problem and for
    valuable discussions. Also I wish to convey my sincerest thanks to
    H. Spruit for the kind hospitality at the Max-Plank Institute for
    Astrophysics, under the auspices of which this work was performed.
    The work in Russia is partly supported by RBRF grant 02-02-16500.
\end{acknowledgements}

\end{document}